\documentclass[showpacs,preprintnumbers,preprint,aps,amsmath,amssymb]{revtex4-1}
\usepackage{color}
\usepackage{graphics}
\usepackage{amsmath,amssymb}
\newcommand{\mathsym}[1]{{}}

\def\10{$SO(10)$}

\newcommand{\ba}{\begin{array}}
\newcommand{\ea}{\end{array}}
\newcommand{\be}{\begin{equation}}
\newcommand{\ee}{\end{equation}}
\newcommand{\beqa}{\begin{eqnarray}}
\newcommand{\eeqa}{\end{eqnarray}}
\def\321{$SU(3)\times SU(2)\times U(1)$}

\newcommand{\dms}  {\Delta m^2_{sol}}
\newcommand{\dma}  {\Delta m^2_{atm}}

\begin{document}
\vspace*{1cm}
\title{An $SO(10)\times S_4$ Model of Quark-Lepton Complementarity}
\bigskip
\author{Ketan M. Patel\footnote{kmpatel@prl.res.in}}
\affiliation{Physical Research Laboratory, Navarangpura, Ahmedabad 380 009, India \vskip 1.0 truecm}

\begin{abstract}
\vskip 0.5 truecm 
The present observations of Cabbibo angle and solar mixing angle satisfy the empirical relation
called Quark-Lepton Complementarity(QLC), namely $\theta_{12}^l\sim \pi/4-\theta_C$. It suggests the
existence of correlation between quarks and leptons which is supported by the idea of grand
unification. We propose a specific ansatz for the structure of Yukawa matrices in $SO(10)$ unified
theory which leads to such relation if neutrinos get masses through type-II seesaw mechanism.
Viability of this ansatz is discussed through detailed analysis of fermion masses and mixing angles
all of which can be explained in a model which uses three Higgs fields transforming as 10 and one
transforming as $\overline{126}$ representations under $SO(10)$. It is shown that the proposed
ansatz can be derived from an extended model based on the two pairs of 16-dimensional vector-like
fermions and an $S_4$ flavor symmetry. The model leads to the lepton mixing matrix that is
dominantly bimaximal with ${\cal O}(\theta_C)$ corrections related to quark mixings. A generic
prediction of the model is the reactor angle $\theta_{13}^l\sim \theta_C/\sqrt{2}$ which is close
to its present experimental upper bound.\\
\end{abstract}
\pacs{11.30.Hv, 12.10.-g, 12.15.Ff, 14.60.Pq}

\maketitle

\section{Introduction}

Experiments on neutrino oscillations have revealed that two of the three leptonic mixing angles are
large. One of them called the atmospheric mixing angle is almost maximal $\theta_{23}^l=
42.3^{o}\left(^{+11.4}_{-7.1} \right)$ and the other known as the solar mixing angle $\theta_{12}^l=
34.5^{o}\left(^{+3.2}_{-2.8} \right)$ is smaller compared to it \cite{garcia}. In contrast, the
observed quark mixing angles are small and hierarchical. The largest angle is the Cabibbo angle
$\theta_{C}\equiv\theta_{us}\approx 13^{o}$ while other two are $\theta_{cb}\approx 2.4^{o}$ and
$\theta_{ub}\approx 0.2^{o}$. An understanding of such wide dissimilarity between the quark and
lepton mixing patterns is considered as one of the major challenges for the physics beyond the
standard model. It has been observed long ago \cite{ qlc:smirnov} that
there exists an interesting empirical relation between quark and lepton mixing
angles.
\beqa \label{qlc}
\theta_{12}^l+\theta_{us} \sim \dfrac{\pi}{4}\eeqa
The above relation is known as Quark-Lepton Complementarity \cite{qlc, minakata, qlcgut, qlcps}
and still favored by the present experimental data within their measurement errors. It is also
possible to write similar relation between 23 angles
of quark and lepton mixing.
\beqa \label{qlc2}
\theta_{23}^l+\theta_{cb} \sim \dfrac{\pi}{4}\eeqa
If such relations are not accidental, they strongly suggest the common roots between quarks and
leptons \cite{ minakata, qlcgut, qlcps}. Clearly it is very hard to realize such relations in
ordinary bottom-up approaches where the quarks and leptons are treated separately with no specific
connections between them. So one requires top-down approaches like the Grand Unified Theories(GUT)
which sometime also unify quarks and leptons and provide a framework to construct a model in which
QLC relation can be embedded in a natural way.\\

The general conditions under which QLC relation (\ref{qlc}) can be realized from quark-lepton
unification are thoroughly discussed in \cite{qlcgut, minakata}. We describe one such possibility
here. The quark mixing matrix known as Cabibbo-Kobayashi-Maskawa(CKM) matrix is defined as
$V_{CKM}=U_u^\dagger U_d$ where $U_u(U_d)$ is unitary matrix that diagonalizes the up-(down-)type
quark mass matrix. Corresponding leptonic mixing matrix, also called
Pontecorvo-Maki-Nakagawa-Sakata(PMNS) matrix, is $V_{PMNS}=U_e^\dagger U_{\nu}$. Assume that the
structure of neutrino and quark mass matrices at high scale are such that the PMNS matrix is exact
bimaximal $V_{PMNS}=U_{BM}$ whereas the CKM matrix is an identity matrix to a leading order. 
\be \label{ubm}
U_{BM}=
\left( \ba{ccc}
\frac{1}{\sqrt{2}}&-\frac{1}{\sqrt{2}}&0\\\frac{1}{2}&\frac{1}{2}&-\frac{1}{
\sqrt{2 }}\\
\frac{1}{2}&\frac{1}{2}&\frac{1}{\sqrt{2}}\\ \ea \right)
\ee
Both the mixing matrices get corrected by ${\cal O}(\theta_C)$ terms coming from the
next leading order where the down quark and charged lepton mass matrices are equal (or nearly so).
In this scenario, a QLC relation can emerge from quark-lepton unification at high scale.
Construction of a realistic GUT model in which all fermion masses and mixing angles are correctly
reproduced along with QLC is highly non-trivial. In fact several models \cite{qlcps} proposed to
explain QLC are based on a smaller gauge group, namely Pati-Salam $SU(4)_c\times SU(2)_L \times
SU(2)_R $ group. A complete and realistic model based on \10 GUT has not been proposed so far. The
original proposal \cite{ qlcgut} was based on $SU(5)$ relation $M_e=M_d^{T}$ but detailed
explanation of the fermionic spectrum was not developed. Here we present a predictive \10 based
unified description of fermion masses and mixing in which QLC relation can be naturally realized.\\

The renormalizable models based on the \10 gauge group are quite powerful in
constraining the fermion mass structure. Moreover, they provide a natural framework for
understanding neutrino masses because of the seesaw mechanisms inherent in them. Fermion masses
arise in these models through their couplings to Higgs fields transforming as ${\bf 10}$,
$\overline{\bf126}$ and ${\bf 120}$ dimensional representations under \10. Neutrino masses
arise from two separate sources either from the vacuum expectation value (vev) of the
right handed triplet (type-I) or from the left handed triplet (type-II) Higgs. The minimal model
with ${\bf 10}$ and $\overline{\bf126}$ Higgs fields has attracted a lot attention \cite{
msgut,btau,typeIIgut}. There also exist a class of models where appropriate flavor symmetry is
integrated with \10 framework with extended Higgs sector \cite{ flavorgut1, flavorgut2,
Dutta:2009bj, Dutta:2009ij} to construct a predictive theory which can simultaneously explain
hierarchical nature of quark masses and mixing angles and large lepton mixing angles. In this paper,
we show that QLC relation follows in a specific \10 model combined with $S_4$ symmetry if dominant
source of neutrino mass is type-II. An additional $Z_n$ symmetry is required in the model to get
desired interactions between various fields.\\

The paper is organized as follows. We describe the fermion mass relations in the model based on
renormalizable supersymmetric (SUSY) \10 GUT in the next section. In section III, we propose a
specific ansatz which predictively interrelates various observables of quark and lepton sectors and
leads to QLC relation. We also discuss the phenomenological implications of such ansatz in this
section. In section IV, we justify the proposed ansatz by a flavor symmetry group $S_4\times Z_n$.
Finally we conclude in section V.\\

\section{Renormalizable SUSY \10 model for fermion masses}

We consider three families of {\bf16}-dimensional fermions obtaining their masses from
renormalizable
couplings to four Higgs multiplates, three of them (denoted by $\Phi, \Phi'$ and $\Phi'' $)
transforming as {\bf10} and the other ($\overline{\Sigma}$) as $\overline{{\bf126}}$ dimensional
representations
under
\10. The \10 breaking can be achieved with ${\bf210}+{\bf54}+{\bf126}+\overline{{\bf126}}$
\cite{typeIIgut}.
The Yukawa
interactions of the model can be written as \cite{flavorgut1}
\beqa \label{yukawa-sup-pot}
W_Y=Y_{10} \psi \psi \Phi +Y_{\overline{126}} \psi \psi \overline{\Sigma} + Y_{10'} \psi \psi
\Phi'+Y_{10''}
\psi \psi \Phi''
\eeqa
where $Y_i$ are symmetric Yukawa coupling matrices. The representations $\Phi, \Phi',
\Phi''$ and $\overline{\Sigma}$ have two minimal supersymmetric standard model (MSSM) doublets in
each of them.
It is assumed that only one linear combination of the up-type doublets and one of the down-type
doublets remain light and play the role of $H_u$ and $H_d$ fields. Once these light doublets acquire
vacuum expectation values, they break electroweak symmetry and generate the fermion masses as
well. The resulting fermion mass matrices can be suitably written as 
\beqa \label{matrices}
M_d&=& H+F+t~H'+H'';\nonumber \\
M_u&=& r~H+s~ F+H'+p~H''; \nonumber \\
M_e&=& H-3 F+t~H'+H'';\nonumber\\
M_D&=& r~H-3 s~ F+H'+p~H'';\nonumber \\
M_L&=& r_L ~F;\nonumber \\
M_R&=& r_R^{-1}~ F.\eeqa
where $H,F,H'$ and $H''$ are obtained by multiplying electroweak vevs and Higgs mixing parameters
with Yukawa coupling matrices $Y_{10}, Y_{\overline{126}}, Y_{10'}$ and $Y_{10''}$ respectively.
$r,s,t,p,r_L$ are $r_R$ are dimensionless parameters determined by the Clebsch-Gordan coefficients,
ratios of vevs, and mixing among Higgs fields (see \cite{flavorgut2} for example). $M_D$ denotes
neutrino Dirac mass matrix. $M_L(M_R)$ is the Majorana mass matrix for left-(right-)handed
neutrinos which receives a contribution only from the vev of $\overline{\Sigma}$ field. In generic
\10 models
of this type, the effective neutrino mass matrix ${\cal M}_\nu$ for the three light neutrinos has
type-I and type-II contributions.
\be \label{mnu}
{\cal M}_\nu\equiv{\cal M}_\nu^{II}+{\cal M}_\nu^I=r_L F-r_R M_D F^{-1} M_D^T. \ee
In general, both contributions are present and they depend on two different parameters so one may
dominate over the other. It has been shown in several references \cite{ typeIIgut} that it is
possible to have symmetry breaking pattern in \10 where type-II term dominates over the type-I
contributions. In this limit, neutrino masses and mixing are governed by $F$ which can be
written as $F\sim M_d-M_e$. It is well known that this relation establish interesting relationship
between $b-\tau$ unification and large atmospheric mixing angle\cite{ btau}. The equations
(\ref{matrices}) and (\ref{mnu}) are the key equations that provide basic platform to construct
a model in which the QLC relation (\ref{qlc}) can be realized.

\section{Ansatz}

We propose following ansatz which leads to relation (\ref{qlc}).

\be \label{ansatz1} 
\ba{ccc}
H=\dfrac{1}{2}\left( \ba{ccc} 0&0&0\\0&h&h\\
0&h&h\\ \ea \right);
&
F=\left( \ba{ccc}
b+c&\sqrt{2}a&0\\\sqrt{2}a&b+c&0\\0&0&b-c\\ \ea
\right);
&
H'=\left( \ba{ccc}
0&0&\sqrt{2}a'\\0&0&0\\\sqrt{2}a'&0&0\\ \ea
\right);
\ea 
H''=x~I\ee
where $I$ is $3\times3$ identity matrix. To do the simple analytical study of such ansatz we assume
that all the above parameters are real. Without loss of generality, we can express the above
matrices in a basis with diagonal $H$. Such basis are obtained by rotating the {\bf16}-dimensional
fermion fields in 2-3 plane by an angle $\pi/4$. The matrices in (\ref{ansatz1}) will be redefined
in new basis as 
\be
(H,F,H',H'')\rightarrow R_{23}\left( \dfrac{\pi}{4}\right) 
(H,F,H',H'')R_{23}^{T}\left(\dfrac{\pi}{4}\right)
\ee
and can be rewritten as

\be \label{ansatz2} 
\ba{ccc}
H=\left( \ba{ccc} 0&0&0\\0&0&0\\
0&0&h\\ \ea \right);
&
F=\left( \ba{ccc}
b+c&a&a\\a&b&c\\a&c&b\\ \ea
\right);
&
H'=\left( \ba{ccc}
0&-a'&a'\\-a'&0&0\\a'&0&0\\ \ea
\right);
\ea 
H''= x~I\ee

Before we present the detailed analysis let us look at some immediate implications of the above
ansatz. The dominant {\bf10}-Higgs coupling matrix
$H$ has rank-1. As it was pointed out in \cite{Dutta:2009ij, Dutta:2009bj} this can simultaneously
explain both the observed hierarchy of quark masses as well as the origin of large lepton mixings if
the light neutrino masses are generated through type-II seesaw mechanism. Assuming only one
{\bf10}-Higgs $H$ contribution in charged fermion mass matrices, we get at zeroth order,
\beqa \label{zerothorder}
m_b=m_{\tau}=\dfrac{1}{r} m_t ;~ V_{CKM} = I; ~ V_{PMNS} = U_{BM}.\eeqa
Correct $b-\tau$ unification and large lepton mixings (bimaximal) are obtained with no mixings
between quarks. The charged fermions of first two generations are massless in this case. Further,
the contributions coming from other Higgs coupling matrices $F, H'$ and $H''$ make the model
realistic by giving nonzero masses to first two fermion generations as well as by perturbing both
the
mixing matrices which reproduce observed mixing patterns for both the quark and lepton sectors.\\

We now present the detailed analysis of ansatz(\ref{ansatz2}). Substituting it in
eq.(\ref{matrices}) and (\ref{mnu}), we get
\beqa \label{massmatrices}
M_u=\left( \ba{ccc} s(b+c)+x'&sa-a'&sa+a'\\sa-a'&sb+x'&sc\\
sa+a'&sc&rh+sb+x'\\ \ea \right)&;&
M_d=\left( \ba{ccc} b+c+x&a-ta'&a+ta'\\a-ta'&b+x&c\\
a+ta'&c&h+b+x\\ \ea \right);\nonumber \\
M_e=\left( \ba{ccc} -3(b+c)+x&-3a-ta'&-3a+ta'\\-3a-ta'&-3b+x&-3c\\
-3a+ta'&-3c&h-3b+x\\ \ea \right)&;&
M_{\nu}=r_L~F
\eeqa
where $x'=px$. Since each mass matrix is real symmetric, it can be
diagonalized by a rotation matrix parameterized (in the standard parameterization) by three angles.
\beqa \label{diag}
R_{f}^{T} M_{f} R_{f}& =& Diag(m_{f1},m_{f2},m_{f3});\\ \nonumber
R_f&=&R_{23}(\theta_{23}^{f}) R_{13}(\theta_{13}^{f}) R_{12}(\theta_{12}^{f})\eeqa
where $f=d,u,e,\nu$ and $R_{ij}$ is a rotation matrix in $ij$ plane. The charged fermion mass
matrices are hierarchical ($h\gg b,c \gg a,a' \gg x,x'$) and can be approximately diagonalized
by Jacobi rotation. The results obtained from such diagonalization for the quark sector are
displayed below.

\beqa \label{downmass}
m_b&\approx& h+b+x+{\cal O}\left( \frac{c^2}{h}\right);\nonumber \\
m_s&\approx& b+x+ \frac{(a-ta')^{2}}{b} \left( 1-\frac{x}{b}\right) +{\cal
O}\left( \frac{c^2}{h}\right); \nonumber \\
m_d&\approx& b+c+x- \frac{(a-ta')^{2}}{b} \left( 1-\frac{x}{b}\right)+{\cal O}\left(
\frac{a^{2}}{h}\right).\\ 
\nonumber \\
m_t&\approx& rh+sb+x'+{\cal O}\left( \frac{s^2 c^2}{rh}\right);\nonumber \\
m_c&\approx& sb+x'+ \frac{(sa-a')^{2}}{sb}\left( 1-\frac{x'}{sb}\right)  +{\cal O}\left(
\frac{s^2 c^2}{rh}\right); \nonumber \\
m_u&\approx& s(b+c)+x'- \frac{(sa-a')^{2}}{sb} \left( 1-\frac{x'}{sb}\right)+{\cal O}\left(
\frac{s^2 a^2}{rh}\right). \eeqa
 
\beqa \label{downangle}
\ba{ccc}
\theta_{12}^{d}\approx -\dfrac{a-ta'}{b}\left(
2+\dfrac{c-x}{b}\right);&~\theta_{23}^{d}\approx -\dfrac{c}{h};&~\theta_{13}^{d}\approx
-\dfrac{a+ta'}{h}\left(
1+\dfrac{c}{h}\right).
\ea
\eeqa
\beqa \label{upangle}
\ba{ccc}
\theta_{12}^{u}\approx -\dfrac{sa-a'}{sb}\left(
2+\dfrac{sc-x'}{sb}\right);&~
\theta_{23}^{u}\approx -\dfrac{sc}{rh};&~
\theta_{13}^{u}\approx -\dfrac{sa+a'}{rh}\left(
1+\dfrac{sc}{rh}\right).
\ea
\eeqa

Let us underline some important points in connection with above relations.
\begin{itemize}
\item The six real parameters $h,b,x,r,s,x'$ can be approximated from the six quark masses. $m_b$
and $m_s$ determine the parameters $h$ and $b$. It is easy to see that $r\approx m_t/m_b$
and $s\approx m_c/m_s$ are required to obtain the masses of heavy quarks $m_t$ and $m_c$.
Further, $m_d$ and $m_u$ fix the values of $x$ and $x'$. Since $b,c \gg x$, we require $c\sim -b$ to
obtain small masses of first generation fermions.

\item Let us assume that $a'\approx sa$ in order to keep $\theta_{12}^{u}
\ll \theta_{12}^{d}$. Also note that $\theta_{23}^{u} \approx (s/r) \theta_{23}^{d} \ll
\theta_{23}^{d}$ and $\theta_{13}^{u} \sim (s/r) \theta_{13}^{d} \ll \theta_{13}^{d}$.
In this limit, the quark mixing matrix takes the form 
\be \label{ckm}
V_{CKM}=U_{u}^{\dagger}U_{d}\approx U_{d} \approx R_{23}(\theta_{23}^{d})
R_{13}(\theta_{13}^{d}) R_{12}(\theta_{12}^{d})
\ee 

\item The elements of the CKM matrix fix some more parameters as follows.
\beqa \label{mixings}
c\sim -V_{cb}~h;~~ a-ta'\sim -V_{us}~b;~~a+ta'\sim -V_{ub}~h. 
\eeqa
An interesting relationship between $V_{us}$ and $V_{ub}$ can be found in the limit
$t\sim0$. 
\beqa \label{1213}
V_{ub}\approx V_{us} \dfrac{m_s}{m_b}+{\cal O}\left( \dfrac{m_s^2}{m_b^2}\right) 
\eeqa 
We will show later in this section that $t\sim0$ is a necessary requirement to obtain QLC
relation(\ref{qlc}).

\item Our assumption of real parameters makes the theory CP invariant. The observed CP violation
in the quark sector can be accommodated by making some parameters complex.\\
\end{itemize}

It is interesting to note that all the parameters are fixed in terms of the observables of the
quark sector. Hence the entire lepton sector emerges as the prediction of
the model. Let us first derive the predictions for the charged leptons.
\beqa \label{clmass}
m_{\tau}&\approx& h-3b+x+{\cal O}\left( \frac{c^2}{h}\right);\nonumber \\
m_{\mu}&\approx& -3b+x- \frac{(3a+ta')^{2}}{3b} \left( 1+\frac{x}{3b}\right) +{\cal
O}\left( \frac{c^2}{h}\right); \nonumber \\
m_e&\approx& -3(b+c)+x+ \frac{(3a+ta')^{2}}{3b} \left( 1+\frac{x}{3b}\right)+{\cal O}\left(
\frac{a^{2}}{h}\right). \eeqa

\beqa \label{clangle}
\ba{ccc}
\theta_{12}^{e}\approx -\dfrac{3a+ta'}{3b}\left(
2+\dfrac{3c+x}{3b}\right);&~
\theta_{23}^{e}\approx \dfrac{3c}{h};&~
\theta_{13}^{e}\approx \dfrac{3a-ta'}{h}\left(
1+\dfrac{c}{h}\right). 
\ea
\eeqa

Noteworthy features of the above relations are the following,
\begin{itemize}
\item It predicts $m_{\tau}\approx m_b$ and $m_{\mu}\approx-3 m_s$.

\item For $b= -c$, $m_e\approx m_d$ which is viable with observed values of $m_e$ and $m_d$
extrapolated at the GUT scale within $3\sigma$ deviations \cite{dasparida}. However for $b\neq-c$,
any desired value of $m_d/m_e$ can be obtained.

\item For $t\sim0$, $\theta_{12}^{e} \approx \theta_{C}$, $\theta_{23}^{e} \approx -3 \theta_{cb}$
and $\theta_{13}^{e} \approx -3 \theta_{ub}$.\\

\end{itemize}

The light neutrino mass matrix in eq.(\ref{massmatrices}) has the most general form
which can be diagonalized by bimaximal matrix $U_{BM}$. The mass eigenvalues are,
\be \label{numass}
m_1 = m_0 (b+c+\sqrt{2} a);~m_2 = m_0 (b+c-\sqrt{2} a);~m_3 = m_0 (b-c)
\ee
Interestingly, for $b=-c$ (which can now also be written as
$V_{cb}\approx m_s/m_b$), we get the partial degenerate neutrino mass spectrum $m_1=-m_2 \ll
m_3$ which leads to vanishing solar (mass)$^2$ difference ($\dms=m_2^2-m_1^2=0$) at high scale. We
performed numerical study and found that the radiative corrections to the original neutrino
mass matrix are unable to generate the required splitting between $m_1$ and $m_2$. 
Another way to induce non zero value of $\dms$ is to
allow type-I contribution to the original type-II seesaw neutrino mass matrix.
However such contribution is highly hierarchical (like $M_t^2$) and it largely contributes to the
33 element of neutrino mass matrix which ultimately spoils the nice symmetry of
neutrino mass matrix and hence the bimaximality of neutrino mixings. This
forces us to consider the case where $V_{cb}\neq m_s/m_b$. In this case
we obtain the following expression for the ratio of the solar to atmospheric squared mass
difference.

\be \label{ratio}
\frac{\dms}{\dma}\approx \sqrt{2} V_{us} \left(
\frac{m_s/m_b}{V_{cb}} \left(1+\frac{m_s}{m_b}\right)- 1 
\right) 
\ee

Note that one requires $m_s/m_b\sim 1.08~V_{cb}$ to obtain the observed
value of $\dms/\dma(\sim 0.031)$ and it implies that $V_{cb} < m_s/m_b$ which is not
favored by their present observed values extrapolated at the GUT scale. However as argued in
\cite{Dutta:2009bj}, the threshold corrections to $b-s$ quark mass mixing from gluino and wino
exchange via one-loop diagrams can give desired value of $V_{cb}$. The required deviation from
$b=-c$ is quantified by
 $$ b+c \approx m_s\left(1-\frac{V_{cb}}{m_s/m_b} \right) \lesssim 0.08~m_s $$
which is small and of order of first generation fermion masses and hence allows the correct $m_d$
in eq.(\ref{downmass}).\\

The leptonic mixing matrix can be seen as dominant bimaximal mixing
resulting from neutrino mass matrix and then corrected by ${\cal O}(\theta_C)$ terms coming
from the unitary matrix $U_e$ which diagonalize charged lepton mass matrix.

\be \label{ckm}
V_{PMNS}\equiv U_{e}^{\dagger}U_{\nu}=U_{e}^{T} U_{BM}
\ee
where $U_{e}=R_{23}(-3 \theta_{cb}) R_{13}(-3 \theta_{ub}) R_{12}(\theta_C)$. The resulting
neutrino mixing parameters are the following. 

\beqa \label{mixing}
U_{e2}&\equiv& (V_{PMNS})_{12} \approx -\frac{1}{\sqrt{2}}+\frac{(V_{us}-3 V_{ub})}{2};\nonumber \\
U_{\mu3}&\equiv& (V_{PMNS})_{23}\approx -\frac{1}{\sqrt{2}}(1+3 V_{cb});\nonumber \\
U_{e3}&\equiv& (V_{PMNS})_{13}\approx-\frac{1}{\sqrt{2}}(V_{us}+3 V_{ub}).
\eeqa

The correction of ${\cal O}(\theta_C)$ from charged lepton generates correct solar mixing angle
which follows QLC relation (\ref{qlc}). The atmospheric mixing angle gets considerable deviation
$\theta_{23}^l\approx\dfrac{\pi}{4}+3\theta_{cb}$ in this
model unlike the standard QLC relation for 23 mixing angle of quark and lepton given in 
eq.(\ref{qlc2}). The model also predicts large value of $U_{e3}\approx0.16$ which can
be tested in planned long baseline experiments.\\ 

Note that eq.(\ref{mixing}) holds at GUT scale which might be changed by RGE corrections in
principle. However it is known that running of the Cabibbo angle is negligibly small in MSSM even
with large value of tan$\beta$. Running of leptonic mixing angle depends on the type of mass
spectrum of light neutrinos. For $b\neq-c$, neutrino mass spectrum follows normal hierarchy
$m_1<m_2\ll m_3$. The effect of RGE corrections are known to be negligible in this case and
eq.(\ref{mixing}) holds at low scale also.\\

We now provide an example of values of the parameters of eq.(\ref{ansatz2})
which successfully generate entire fermion mass spectrum as well as mixing patterns for both
quark and lepton sector. The required CP violation in the quark sector is incorporated by making
$a'$ complex. In the limit $t\sim0$, $a'$ contributes only to the up quark mass matrix and does not
change the other predictions of ansatz given in eq.(\ref{ansatz2}). One more parameter $x'$ is made
complex to reproduce $m_u$ correctly. The numerical values of parameters are
\beqa \label{prm}
h&=&1.7~{\rm GeV};~b=0.0243~{\rm GeV};~c=-0.022113~{\rm GeV};~a=-0.0052~{\rm GeV};\nonumber \\
a'&=&(0.0344247-0.028885i)~{\rm GeV};~x'=(0.0233596-0.00293374i)~{\rm GeV};\nonumber \\
x&=&0.00325~{\rm GeV};~r=55.88;~s=-8.64198;~t=0. \eeqa

Substituting these numbers in eq.(\ref{massmatrices}), we get
\beqa \label{resmass}
\ba{ccc}
m_t=94.8~{\rm GeV};&~ m_c=0.19~{\rm GeV};&~ m_u=0.65~{\rm MeV};\\
m_b=1.73~{\rm GeV};&~ m_s=28.5~{\rm MeV};&~ m_d=4.21~{\rm MeV}; \\
m_{\tau}=1.63~{\rm GeV};&~ m_{\mu}=75.4~{\rm MeV};&~ m_e=0.35~{\rm MeV}.
\ea
\eeqa
 
\beqa \label{resmixings}
\ba{cccc}
{\rm sin}\theta_{us}=0.222;&~ {\rm sin}\theta_{cb}=0.015;&~ {\rm sin}\theta_{ub}=0.005;&~
\delta_{CKM}= 60.9^{\circ};\\
{\rm sin}^2 \theta^l_{12}=0.368;&~ {\rm sin}^2 \theta^l_{23}=0.527;&~ {\rm sin}^2
\theta^l_{13}=0.024;&~ \frac{\dms}{\dma}=0.030.\\ \ea \eeqa

The obtained spectrum is in good agreement with the data extrapolated at the GUT
scale. For example, we compare our results with the charged fermion masses obtained at the GUT scale
in the MSSM for tan$\beta$=55, $M_{SUSY}=1~{\rm TeV}$ and $M_{GUT}=2\times10^{16}~{\rm GeV}$ given
in table 5 of reference \cite{ dasparida}. All charged fermion masses (except $m_d$) obtained here
fits with the data within $1\sigma$. Our ansatz predicts larger value of $m_d$. The quark mixing
angles $\theta_{cb}$ is small ($<m_s/m_b$) as required by eq.(\ref{ratio}). The reproduced values of
lepton mixing angles and $\dms/\dma$ are also in accordance with their updated low energy values
(within 3$\sigma$ measurement errors) given in \cite{garcia}.\\

\section{The Model}

In this section, we will illustrate how the ansatz (\ref{ansatz1}) can be obtained in a model from 
flavor symmetry. We use discrete flavor symmetry based on the group $S_4$ which is a group of
permutation of four distinct objects. It has 24 distinct elements filled in five conjugate classes
and hence five irreducible representations of
dimensions ${\bf3}_{\bf2},{\bf3}_{\bf1},\bf{2},{\bf1}_{\bf2}$ and ${\bf1}_{\bf1}$. A singlet
representation with subscript ``2'' changes sign under transformation involving the odd number of
permutations of $S_4$. More details on the group theory of $S_4$, its multiplication rules and
the Clebsch-Gordan coefficients are reported in \cite{lindner}.\\

Our model follows the same line as model constructed in \cite{Dutta:2009bj} and
uses the same symmetry group. However it differs at some places since the ansatz required here is
different from their ansatz. The basic matter fields and Higgs fields content of the model is
the same as discussed in section II. In addition to
this we use five flavon fields which are singlets under \10 and two pair of vector-like fermion
fields
which transform like ${\bf16}\oplus\overline{{\bf16}}$ under \10. We impose the $S_4$ symmetry
together with
$Z_n$ symmetry to get desired structure of Yukawa matrices. Three matter fields $\psi$ are assigned
as ${\bf3}_{\bf2}$ dimensional representation of $S_4$ while five flavon fields $\chi,\phi,
\eta, \sigma$ and
$\sigma'$ form ${\bf3}_{\bf1},{\bf3}_{\bf2},{\bf3}_{\bf1},{\bf1}_{\bf1}$ and ${\bf1}_{\bf2}$
representations of $S_4$ respectively. The other fields are
singlet (${\bf1}_{\bf1}$ or ${\bf1}_{\bf2}$) under $S_4$. An additional $Z_n$ symmetry is required
to allow/forbid
interactions between particular fields. The $Z_n$ charges of various fields are listed in
table(\ref{tab:1}) where $\omega = e^{i(2\pi/n)}$.\\
 
\begin{table} [h]
\begin{small}
\begin{math}
 \begin{array}{|c|c|cccc|ccccc|cccc|}
\hline
 & \psi & \Phi & \Phi' & \Phi'' & \overline{\Sigma} & \chi & \phi & \eta
& \sigma & \sigma'  & \Psi_{V1} & \overline{\Psi}_{V1} & \Psi_{V2} & \overline{\Psi}_{V2}\\
\hline
\10 & {\bf16} & {\bf10} & {\bf10} & {\bf10} & \overline{{\bf126}} & {\bf1} & {\bf1} & {\bf1} &
{\bf1} & {\bf1} & {\bf16} & \overline{{\bf16}} & {\bf16}
&\overline{{\bf16}}\\
\hline
S_4 & {\bf3}_{\bf2} & {\bf1}_{\bf1} & {\bf1}_{\bf2} & {\bf1}_{\bf1} & {\bf1}_{\bf1} & {\bf3}_{\bf1}
& {\bf3}_{\bf2} & {\bf3}_{\bf1} & {\bf1}_{\bf1} & {\bf1}_{\bf2} & {\bf1}_{\bf1} & {\bf1}_{\bf1} &
{\bf1}_{\bf2} & {\bf1}_{\bf2}\\
\hline
Z_n & 1 & \omega^{-2m} & \omega^{-(p+q)} & \omega^{-2q} & \omega^{-2k} & \omega^{k} & \omega^{m} &
\omega^{p} & \omega^{k} & \omega^{q} & \omega^{m} & \omega^{-m} & \omega^{k} & \omega^{-k}\\
\hline
\end{array}
\end{math}
\end{small}
\caption{Various fields and their representations under $SO(10) \times\ S_4 \times Z_n$.}
\label{tab:1}
\end{table}

Let us consider a theory above GUT scale which is invariant under the symmetry group $SO(10) \times
S_4 \times Z_n$. The Yukawa superpotential allowed by such symmetry can be written
as

\beqa \label{suppot}
W&=& (\phi \psi) \overline{\Psi}_{V1}+\lambda \Psi_{V1} \Psi_{V1} \Phi + M_1 \Psi_{V1}
\overline{\Psi}_{V1}\nonumber \\
&+&(\chi \psi) \overline{\Psi}_{V2}+\lambda' \Psi_{V2} \Psi_{V2} \overline{\Sigma} + M_2 \Psi_{V2}
\overline{\Psi}_{V2}\nonumber \\
&+&\sum_i \dfrac{\alpha_i}{\Lambda^2} (\chi^2 \psi \psi)_i \overline{\Sigma}
+\dfrac{\beta}{\Lambda^2} 
\sigma (\chi \psi \psi) \overline{\Sigma} +\dfrac{\gamma}{\Lambda^2} \sigma^2 (\psi \psi)  \overline{\Sigma} \nonumber \\
&+&\dfrac{\alpha'}{\Lambda^2} \sigma' (\eta \psi \psi) \Phi'
+\dfrac{\alpha''}{\Lambda^2} \sigma'^2 (\psi \psi) \Phi''\eeqa
where $\Lambda$ is the Planck scale up to which the theory is valid. The $S_4$
singlet contraction of flavor index is indicated with bracket. $\alpha_i,\alpha', \alpha'', \beta,
\gamma, \lambda$, and $\lambda'$ are coefficients of $\cal{O}$(1). The term $(\chi^2 \psi \psi)$
represents all the different $S_4$ contractions which can be constructed as follows:

\beqa \label{quartic}
(\chi^2 \psi \psi)_i &\equiv& ((\chi \chi)_{{\bf1}_{\bf1}}(\psi
\psi)_{{\bf1}_{\bf1}}),((\chi \chi)_{\bf2}(\psi \psi)_{\bf2}),((\chi \chi)_{{\bf3}_{\bf1}}(\psi
\psi)_{{\bf3}_{\bf1}}),((\chi \chi)_{{\bf3}_{\bf2}}(\psi \psi)_{{\bf3}_{\bf2}}),\nonumber \\
& &((\chi \psi)_{{\bf1}_{\bf2}}(\chi \psi)_{{\bf1}_{\bf2}}),((\chi \psi)_{\bf2}(\chi
\psi)_{\bf2}),((\chi \psi)_{{\bf3}_{\bf1}}(\chi \psi)_{{\bf3}_{\bf1}}),((\chi
\psi)_{{\bf3}_{\bf2}}(\chi \psi)_{{\bf3}_{\bf2}}) \eeqa
where $(...)_{R}$ indicates the representation $R$ under $S_4$. Now consider a theory
below the scale of $M_{1,2}$ and at the GUT scale. The effective superpotential after integrating
out heavy vector-like fields is given by,
\beqa \label{effsuppot}
W_{eff}&=& \dfrac{\lambda}{M_1^2} (\phi \psi)(\phi \psi)\Phi + \dfrac{\lambda'}{M_2^2} (\chi
\psi)(\chi \psi)\overline{\Sigma} \nonumber \\
&+&\sum_i \dfrac{\alpha_i}{\Lambda^2} (\chi^2 \psi \psi)_i \overline{\Sigma}
+\dfrac{\beta}{\Lambda^2} 
\sigma (\chi \psi \psi) \overline{\Sigma} +\dfrac{\gamma}{\Lambda^2} \sigma^2 (\psi \psi)  \overline{\Sigma} \nonumber \\
&+&\dfrac{\alpha'}{\Lambda^2} \sigma' (\eta \psi \psi) \Phi'
+\dfrac{\alpha''}{\Lambda^2} \sigma'^2 (\psi \psi) \Phi''\eeqa
where first two terms allow the desired rank-1 structure of Yukawa matrices. Note that effective
Yukawa superpotential still has the symmetry $SO(10) \times S_4 \times Z_n$. This symmetry will
be
broken to \10 by vevs of the flavon fields. In order to get the desired structure of Yukawa
couplings,
we will choose particular vacuum alignment of the flavon fields as given below.
\be \label{flavonvevs} 
\ba{ccccc}
\langle\phi\rangle=\left( \ba{c} 0\\1\\
1\\ \ea \right) \upsilon_{\phi};~
&
\langle\chi\rangle=\left( \ba{c} 0\\0\\
1\\ \ea \right) \upsilon_{\chi};~
&
\langle\eta\rangle=\left( \ba{c} 0\\1\\
0\\ \ea \right) \upsilon_{\eta};~
&
\langle\sigma\rangle=\upsilon_{\sigma};~
&
\langle\sigma'\rangle=\upsilon_{\sigma'}\\
\ea 
\ee
These vevs of flavon fields break flavor symmetry $S_4$ at the GUT scale and generate following
structure of various Yukawa couplings.
\beqa \label{yukawa10}
Y_{10}=\dfrac{\lambda \upsilon_{\phi}^2}{M_1^2}\left( \ba{ccc}
0&0&0\\0&1&1\\
0&1&1\\ \ea \right)\eeqa
\beqa \label{yukawa126}
Y_{126}=\dfrac{\lambda '\upsilon_{\chi}^2}{M_2^2}\left( \ba{ccc}
0&0&0\\0&0&0\\
0&0&1\\ \ea \right) + \dfrac{\upsilon_{\chi}^2}{\Lambda^2}\left( \ba{ccc}
\tilde{\alpha}&0&0\\0& \tilde{\alpha}&0\\
0&0&\tilde{\alpha_0}\\ \ea \right) +  \dfrac{\beta\upsilon_{\chi} \upsilon_{\sigma}
}{\Lambda^2}\left( \ba{ccc}
0&1&0\\1&0&0\\
0&0&0\\ \ea \right) + \dfrac{\gamma\upsilon_{\sigma}^2 }{\Lambda^2} I \eeqa
\beqa \label{yukawa10'}
Y_{10'}=\dfrac{\alpha' \upsilon_{\sigma'} \upsilon_{\eta}}{\Lambda^2}\left( \ba{ccc}
0&0&1\\0&0&0\\
1&0&0\\ \ea \right)\eeqa
\beqa \label{yukawa10''}
Y_{10''}=\dfrac{\alpha'' \upsilon_{\sigma'}^2 }{\Lambda^2}  I\eeqa
where all non-relevant Clebsch-Gordan coefficients are suitably absorbed.
$\tilde{\alpha}$ and $\tilde{\alpha_o}$ are linear combinations of different $\alpha_i$. 
The Yukawa matrices derived from the super potential can successfully explain the
ansatz given in eq.(\ref{ansatz1}). Note that $M_{2} \ll \Lambda$ which implies $b+c \ll b-c$ (or
$b\approx-c$) in eq.(\ref{ansatz1}). Further, the assumption $M_{1} \ll M_{2}$ leads to $h \gg
b,c$.\\

It is very important to show that the required vacuum structure of flavon fields
(\ref{flavonvevs}) is allowed by flavon superpotential. This point has already been discussed in
great details in reference \cite{Dutta:2009bj}. Since our model has the same kind of flavon
structure as theirs, we simply use their results. Note that due to non-trivial $Z_n$ charges,
bilinear terms which correspond to masses of flavon fields are not allowed. As a result of this the
model requires doubling of flavon fields to allow Dirac type mass terms. The new flavon fields have
the same $S_4$ representations but opposite $Z_n$ charges. It has been shown in \cite{Dutta:2009bj}
that all the desired vacua of eq.(\ref{flavonvevs}) are present in the model.\\

\section{Summary}

In this paper, we have studied a possible way to realize QLC
relation (\ref{qlc}) between the Cabibbo angle and solar mixing angle in realistic quark-lepton
unification theory based on \10 gauge group. We have shown here that it is indeed possible to
obtain such relation starting from the fermionic mass structure (\ref{matrices}) if they are
supplemented with ansatz (\ref{ansatz1}) and assuming that only type-II seesaw mechanism is
responsible for light neutrino masses. One necessary ingredient for QLC is bimaximal mixing
pattern from the neutrino sector which has been obtained through specific ansatz. Our ansatz also
makes use of recently proposed \cite{Dutta:2009ij} rank-1 strategy which naturally explains charged
fermions mass
hierarchy as well as origin of hierarchical quark mixing angles as opposed to the large lepton
mixing angles. We have shown through the detailed analysis that this ansatz is capable
of explaining the entire fermionic spectrum and not just the QLC relation. Moreover, the
various predictions made by such ansatz are in agreement with observations. We have shown that
the proposed ansatz can be obtained in a model from a discrete flavor symmetry group $S_4$ together
with an additional $Z_n$ symmetry. A generic prediction of our approach is $\theta_{13}^l\approx
\theta_C/\sqrt{2}$ which is near to its current experimental upper bound. The atmospheric mixing
angle gets considerable deviation from maximality ($\theta_{23}^l\approx \pi/4+3 \theta_{cb}$) in
this approach. These predictions can be confirmed or excluded by the current generation of neutrino
oscillations experiments.\\
\\
\\
{\bf Acknowledgements}\\
I thank Anjan S. Joshipura for many useful discussions, suggestions and encouragement during this
work and for reading the manuscript carefully.\\

\end{document}